\documentclass[aps,pre,preprint,showpacs]{revtex4}
\usepackage{epsf}
\usepackage{amsmath,amssymb}
\usepackage{times}

\begin{document}

\title {Quantum dephasing and decay of classical correlation
functions in chaotic systems}
\author{Valentin V. Sokolov}
\affiliation{Center for Nonlinear and Complex Systems,
Universit\`a degli Studi dell'Insubria, Via Valleggio 11,
22100 Como, Italy} \affiliation{Budker Institute of Nuclear
Physics, Novosibirsk, Russia}
\author{Giuliano Benenti}
\affiliation{Center for Nonlinear and Complex Systems, Universit\`a
degli Studi dell'Insubria, Via Valleggio 11, 22100 Como, Italy}
\affiliation{CNISM  and Istituto Nazionale di Fisica Nucleare,
Sezione di Milano}
\author{Giulio Casati}
\affiliation{Center for Nonlinear and Complex Systems, Universit\`a
degli Studi dell'Insubria, Via Valleggio 11, 22100 Como, Italy}
\affiliation{CNISM and Istituto Nazionale di Fisica Nucleare,
Sezione di Milano} \affiliation{Department of Physics, National
University of Singapore, Singapore 117542, Republic of Singapore}
\date{\today}
\pacs{05.45.Mt, 03.65.Sq, 05.45.Pq}

\begin{abstract}
We discuss the dephasing induced by the internal classical chaotic
motion in the absence of any external environment. To this end we
consider a suitable extension of fidelity for mixed states which is
measurable in a Ramsey interferometry experiment. We then relate the
dephasing to the decay of this quantity which, in the semiclassical
limit, is expressed in terms of an appropriate classical correlation
function. Our results are derived analytically for the example of a
nonlinear driven oscillator and then numerically confirmed for the
kicked rotor model.
\end{abstract}
\maketitle

\section{Introduction}
\label{sec:intro} The study of the quantum manifestations of
classical chaotic motion has greatly improved our understanding of
quantum mechanics in relation to the properties of eigenvalues,
eigenfunctions as well as to the time evolution of complex systems
\cite{haakebook,ccbook}. According to the Van Vleck-Gutzwiller's
semiclassical theory \cite{Gutzwiller}, the quantum dynamics, even
deeply in the semiclassical region, involves quantum interference of
contributions from a large number of classical trajectories which
exponentially grows with the energy or, alternatively, with time.
This interference manifests itself in various physical effects such
as  universal local spectral fluctuations, scars in the eigenstates,
elastic enhancement in chaotic resonance scattering, weak
localization in transport phenomena and, also, in peculiarities of
the wave packet dynamics and in the decay of the quantum Loschmidt
echo (fidelity) \cite{Peres84}:
\begin{equation}\label{def}
F_{\overset{\circ}\psi}(t)=|\langle \overset{\circ}\psi|{\hat f(t)}|\overset{\circ}\psi\rangle|^2=
\left|Tr\left[{\hat f(t)}\overset{\circ}\rho\right]\right|^2
\end{equation}
In Eq.~(\ref{def}),
$\overset{\circ}\rho=|\overset{\circ}\psi\rangle\langle
\overset{\circ}\psi|$ is the density matrix corresponding to the
initial pure state $|\overset{\circ}\psi\rangle\equiv
|\psi(t=0)\rangle$. The unitary operators ${\hat U}_{0}(t)$ and
${\hat U}_{\varepsilon}(t)$ describe the unperturbed and perturbed
evolutions of the system, according to the Hamiltonians $H_0$ and
$H_{\varepsilon}=H_0+\varepsilon V$, respectively. Therefore, the
echo operator $\hat{f}(t)=\hat{U}_0^\dag(t)\hat{U}_\varepsilon(t)$
represents the composition of a slightly perturbed Hamiltonian
evolution with an unperturbed time-reversed Hamiltonian evolution.
The unperturbed part of the evolution can be perfectly excluded by
making use of the interaction representation, thus obtaining
\begin{equation}\label{fint}
{\hat f(t)}=
T\exp\left[-i\frac{\varepsilon}{\hbar}\int_0^t d\tau {\cal H}(\tau)\right]\, ; \quad {\cal H}(\tau)=e^{\frac{i}{\hbar}H_0\tau}Ve^{-\frac{i}{\hbar}H_0\tau}\,.
\end{equation}
Therefore, the fidelity (\ref{def}) can be seen as the probability, for a system which evolves in accordance with the time-dependent Hamiltonian ${\cal H}(t)=U_0^{\dag}(t) V U_0(t)$, to stay in the initial state $|\overset{\circ}\psi\rangle$ till the time $t$.

The quantity (\ref{def}), whose behavior depends on the interference
of two wave packets evolving in a slightly different way,
measures the stability of quantum motion under perturbations.
Its decay has been investigated extensively in different parameter regimes
and in relation to the nature of the corresponding classical motion
(see  \cite{Peres84,Jalabert01,Jacquod01,Cerruti02,Benenti02,Prosen02,Silvestrov02,
Vanicek03,Cucchietti03,Wang04,eckhardt,Benenti03} and references therein).
 Most remarkably, it turns out that a moderately weak coupling to a disordered
 environment, which destroys the quantum phase correlations thus inducing
 decoherence, yields an exponential decay of fidelity, with a rate
 which is determined by the system's Lyapunov exponent and independent
 of the perturbation (coupling) strength  \cite{Jalabert01}. In other words,
 quantum interference becomes irrelevant and the decay of fidelity is determined
 by classical chaos.
 This result raises the interesting question whether the classical
 chaos, in the absence of any environment and only with a perfectly deterministic
 perturbation $V$, can by itself produce mixing of the quantum phases (dephasing)
 strongly enough to fully suppress the quantum interference. The answer is, generally negative.
  Indeed, though the "effective" Hamiltonian evolution
   (see Eq.~(\ref{fint}))
   is in accordance with the chaotic dynamics of the  unperturbed system so that
    the actions along distant classical phase trajectories are statistically
    independent, still there always exist a lot of very close trajectories whose actions differ only
    by terms of the order of Planck's constant. Interference of such trajectories
    remains strong. In this paper we show that, nevertheless, if the evolution starts
    from a wide and incoherent mixed state, the initial dephasing
    persists due to the intrinsic classical chaos so that quantum phases
    remain irrelevant. We remark in this connection that any classical device is capable of
    preparing only incoherent mixed states described by diagonal density matrices.

Our paper is structured as follows. In Sec.~\ref{sec:mixed}, we
discuss two different definitions of fidelity for mixed states.
Sec.~\ref{sec:iontraps} introduces the kicked nonlinear oscillator
which can serve as a model for an ion trap. The fidelity decay for
this model in the chaotic regime is discussed in
Sec.~\ref{sec:fidpure} (for pure coherent states) and
\ref{sec:fidmixed} (for mixed states). This latter section
analytically establishes a link between dephasing and decay of a
suitable classical correlation function. This link is numerically
confirmed in Sec.~\ref{sec:krot} for the kicked rotor model, whose
fidelity decay may be measured by means of cold atoms in an optical
lattice. Finally, our conclusions are drawn in
Sec.~\ref{sec:conclusions}.

\section{Mixed state fidelities}
\label{sec:mixed}
In the case of a mixed initial state ($\overset{\circ}\rho=
\sum_k p_k|\overset{\circ}\psi_k\rangle\langle\overset{\circ}\psi_k|,\,\, \sum_k p_k=1$), fidelity is usually defined
as \cite{Peres84}
\begin{equation}\label{Fm}
F(t)= \frac{1}{{\rm Tr}(\overset{\circ}\rho^{\,2})} {\rm
Tr}\left[\rho_0(t)\rho_{\varepsilon}(t)\right]=
\frac{1}{{\rm Tr}(\overset{\circ}\rho^{\,2})} {\rm
Tr}[{\hat f^{\dag}(t)}\overset{\circ}\rho{\hat
f(t)}\overset{\circ}\rho].
\end{equation}
Note that for a pure state ($\rho^2=\rho\,,p_k=\delta_{k\overset{\circ}k}$)  eq. (\ref{Fm}) reduces to (\ref{def}).

Another interesting possibility is suggested by the experimental configuration with periodically kicked
ion traps proposed in \cite{zoller}. In such Ramsey type interferometry experiments one directly accesses
the fidelity amplitudes (see \cite{zoller}) rather than their square moduli. Motivated by this consideration,
 we analyze in this paper the following natural generalization of fidelity:
\begin{equation}\label{F1}
{\cal F}(t)=\Big|\sum_k p_k
f_k\Big|^2=\sum_k p^2_k F_k(t)+\sum_{k,k'}(1-\delta_{kk'}) p_k p_{k'}f_k(t)f_{k'}^*(t)\,,
\end{equation}
which is obtained by directly extending  formula (\ref{def}) to the
case of an arbitrary mixed initial states $\overset{\circ}\rho$. The
first term in the r.h.s. is the sum of  fidelities $F_k=|f_k|^2$ of
the individual pure initial states with weights $p_k^2$, while the
second, interference term depends on the relative phases of fidelity
amplitudes. If the number $K$ of pure states
$|\overset{\circ}\psi_{k}\rangle$ which form the initial mixed state
is large, $K\gg 1$, so that $p_k\backsimeq 1/K$ for $k\leqslant K$
and zero otherwise, then the first term is $\backsim 1/K$ at the
initial moment $t=0$ while the second term $\backsim 1$. Therefore,
in the case of a wide mixture, the decay of the function ${\cal
F}(t)$ is determined by the interference terms.

The fidelity ${\cal F}(t)$ (Eq.~\ref{F1}) is different from the
mixed-state fidelity $F(t)$ (Eq.~\ref{Fm}) and from the
 {\it incoherent} sum
\begin{equation}\label{incoh}
\overline{F(t)}=\sum_k p_k|f_k|^2
\end{equation}
of pure-state fidelities  typically considered in the literature.
The two latter quantities both contain only transition probabilities
$W_{k k'}=|\langle\overset{\circ}\psi_k|{\hat
f}(t)|\overset{\circ}\psi_k\rangle |^2$. In contrast, the function
${\cal F}$ directly {\it accounts for the quantum interference} and
can be expected to retain quantal features even in the deep
semiclassical region.

Notice that ${\cal F}(t)$ is the quantity naturally measured in
experiments performed on cold atoms in optical lattices \cite{darcy}
and in atom optics billiard \cite{davidson} and proposed for
superconducting nanocircuits \cite{pisa}. This quantity is
reconstructed after averaging the amplitudes over several
experimental runs (or many atoms). Each run may differ from the
previous one in the external noise realization and/or in the initial
conditions drawn, for instance, from a thermal distribution
\cite{davidson}. Note that the averaged (over noise) fidelity
amplitude can exhibit rather different behavior with respect to the
averaged fidelity \cite{pisa}.

\section{Fidelity decay in ion traps: the driven nonlinear oscillator model}
\label{sec:iontraps} As a model for a single ion trapped in an
anharmonic potential we consider a quartic oscillator driven by a
linear multimode periodic force $g(t)$,
\begin{equation}\label{H_0}
H_0=\hbar\omega_0 n+\hbar^2
n^2-\sqrt{\hbar}(a+a^{\dag})g(t)\,,\quad n=a^{\dag}a,\, [a,a^{\dag}]=1\,.
\end{equation}
In our units, the time and parameters $\hbar, \omega_0$ as well as
the strength of the driving force are dimensionless. The period of
the driving force is set to one. We use below the basis of coherent
states $|\alpha\rangle$ which minimize in the semiclassical domain
the action-angle uncertainty relation. These states are fixed by the
eigenvalue problem
$a|\alpha\rangle=\frac{\alpha}{\sqrt{\hbar}}|\alpha\rangle$, where
$\alpha$ is a complex number which does not depend on $\hbar$. The
corresponding normalized phase density (Wigner function) is
$\rho_{\overset{\circ}\alpha}(\alpha^*,\alpha)=
\frac{2}{\pi\hbar}e^{-2\frac{|\alpha-\overset{\circ}\alpha|^2}{\hbar}}$
and occupies a cell of volume $\sim\hbar$ in the phase plane
$(\alpha^*,\alpha)$. Coherent interfering contributions in Van
Vleck-Gutzwiller semiclassical theory just come from the "shadowing"
trajectories which originate from such phase regions. In the
classical limit ($\hbar\rightarrow 0$) the above phase density
reduces to Dirac's $\delta$-function, thus fixing a unique classical
trajectory.

Since the scalar product of two coherent states equals
$\langle\alpha'|\alpha\rangle=\exp\left(-\frac{|\alpha'-\alpha|^2}{2\hbar}+ \frac{i}{\hbar}{\rm Im}({\alpha'}^*\alpha)\right)$, they become orthogonal in the classical limit $\hbar\rightarrow 0$. The Hamiltonian matrix $\langle\alpha'|H_0|\alpha\rangle$ is diagonal in this limit and reduces to the classical Hamiltonian function $H_0^{(c)}=\omega_0|\alpha_c|^2+|\alpha_c|^4-(\alpha_c^*+\alpha_c)g(t)$.  The complex variables $\alpha_c, i{\alpha_c}^*$ are canonically conjugated and are  related to the classical action-angle variables $I_c, \theta_c$ via $\alpha_c=\sqrt{I_c}e^{-i\theta_c},\alpha_c^*=\sqrt{I_c}e^{i\theta_c}$.
The action satisfies the nonlinear integral equation
\begin{equation}\label{I_c}
I_c(t)=\left|{\overset{\circ}\alpha}_c+i\int_0^t d\tau
g(\tau)e^{i\varphi_c(\tau)}\right|^2\equiv |a_c(t)|^2,
\end{equation}
where $\alpha_c(t)=a_c(t)\,e^{-i\varphi_c(t)}$ and
$\varphi_c(t)=\int_0^t d\tau[\omega_0+2I_c(\tau)]$. We assume below that
the initial conditions are isotropically distributed with a density
${\cal P}_{\overset{\circ}\alpha_c}(\overset{\circ}\alpha^*,
\overset{\circ}\alpha)={\cal P}(|\overset{\circ}\alpha-
\overset{\circ}\alpha_c|^2)$ around a fixed phase point
$\overset{\circ}\alpha_c\,$.

 When the strength of the driving force exceeds some critical value, the classical motion becomes chaotic,
  the phase $\varphi_c(t)$ gets random so that its autocorrelation function decays exponentially with time:
\begin{equation}\label{corrphy_c}
\Big|\int d^2{\overset{\circ}{\alpha}} {\cal
P}_{\overset{\circ}\alpha_c}(\overset{\circ}{\alpha}^*,
\overset{\circ}{\alpha})\,
e^{i\left[\varphi_c(t)-\varphi_c(0)\right]}\Big|^2=
\exp\left(-t/\tau_c\right)\,.
\end{equation}
 Moreover, Eqs.~(\ref{I_c}, \ref{corrphy_c}) yield the diffusive growth $<~I_c(t)>={\overset{\circ}I}_c+Dt$ of the mean action,
where ${\overset{\circ}I}_c=<~I_c(t=0)>$. By numerical integration
of the classical equations of motion we have verified this statement
as well as the exponential decay (\ref{corrphy_c}).

\section{Semiclassical evolution and fidelity decay for coherent states}
\label{sec:fidpure}
In what follows, we analytically evaluate both the fidelity $F_{\overset{\circ}\alpha}(t)$ for a pure
coherent quantum state $|\overset{\circ}\alpha\rangle$ (this section) as well as the function ${\cal F}(t)$ for
an incoherent mixed state (Sec.~\ref{sec:fidmixed}) by treating the unperturbed motion semiclassically.
Our semiclassical approach allows us to compute these quantities even for quantally strong
perturbations $\sigma=\varepsilon/\hbar\gtrsim 1$.

The semiclassical evolution
$|\psi_{\overset{\circ}\alpha}(t)\rangle= {\hat
U}_0(t)|{\overset{\circ}\alpha}\rangle$ of an initial coherent state
when the classical motion is chaotic has been investigated in
\cite{Sokolov84}. With the help of Fourier transformation one can
linearize the chronological exponent ${\hat U}_0(t)$ with respect to
the operator $n$ thus arriving at the following Feynman's
path-integral representation in the phase space:
\begin{equation}\label{Psi(t)}
\begin{array}{c}
|\psi_{\overset{\circ}\alpha}(t)\rangle=\int\prod_{\tau}
\frac{d\lambda(\tau)}{\sqrt{4\pi
i\hbar}} \exp\left\{\frac{i}{4\hbar}\int_0^t
d\tau\lambda^2(\tau)-\frac{i}{\hbar}{\rm Im}
[\beta_{\lambda}(t)]\right\}|\alpha_{\lambda}(t)\rangle\;.
\end{array}
\end{equation}
The functions with the subscript $\lambda$ are obtained by
substituting $2I_c\Rightarrow\lambda$ in the corresponding classical
functions:
$\alpha_{\lambda}(t)=\left[{\overset{\circ}\alpha}+i\int_0^t d\tau
g(\tau)e^{i\varphi_{\lambda}(\tau)}\right]e^{-i\varphi_{\lambda}(t)}$
and $\beta_{\lambda}(t)=-i\int_0^t d\tau
g(\tau)\alpha_{\lambda}(\tau)$, where $\varphi_{\lambda}(t)=\int_0^t
d\tau[\omega_0+\lambda(\tau)]$.

As an example we choose a perturbation $V$ which corresponds to a
small, time-independent variation of the linear frequency:
$\omega_0\to \omega_0+\varepsilon$ \cite{Iomin04}. For convenience,
we consider the symmetric fidelity operator: ${\hat f}(t)={\hat
U}_{(+)}^{\dag}(t){\hat U}_{(-)}(t)$, where the evolution operators
${\hat U}_{(\pm)}(t)$ correspond to the Hamiltonians
$H_{(\pm)}=H_0\pm\frac{1}{2}\varepsilon n$, respectively. Using
Eq.~(\ref{Psi(t)}) we express $f_{\overset{\circ}\alpha}(t)$ as a
double path integral over $\lambda_1$ and $\lambda_2$. A linear
change of variables
$\lambda_{1}(\tau)=2\mu(\tau)-\frac{1}{2}\hbar\nu(\tau)$,
$\lambda_{2}(\tau)=2\mu(\tau)+\frac{1}{2}\hbar\nu(\tau)$ entirely
eliminates the Planck's constant from the integration measure. After making the
shift $\nu(t)\rightarrow\nu(t)-\varepsilon/\hbar$ we obtain
$$\begin{array}{c}
f_{\overset{\circ}\alpha}(t)=\int\prod_{\tau}
\frac{d\mu(\tau)d\nu(\tau)}{2\pi}\,\exp\left\{i\sigma\int_0^t
d\tau\mu(\tau) \right. \\\left. -i\int_0^t
d\tau\mu(\tau)\nu(\tau) +\frac{i}{\hbar}{\cal
J}\left[\mu(\tau),\nu(\tau)\right] -\frac{1}{2\hbar}{\cal
R}\left[\mu(\tau),\nu(\tau)\right] \right\},
\end{array}$$
where the functionals ${\cal J}, {\cal R}$ equal
\begin{equation}\label{JR}
\begin{array}{c}
{\cal J}=\hbar\int_0^t d\tau\nu(\tau)|a_{\mu}(\tau)|^2+O(\hbar^3), \\
{\cal R}=\hbar^2|\int_0^t d\tau\nu(\tau)a_{\mu}(\tau)|^2 +O(\hbar^4), \\
\end{array}
\end{equation}
and vanish in the limit $\hbar=0$. The quantities with the subscript $\mu$ are obtained by setting $\nu(\tau)\equiv 0$ (in particular,
$a_\mu(t)=\alpha_\mu(t)e^{i\varphi_\mu(t)}$, with $\alpha_\mu(t)=\left[{\overset{\circ}\alpha}+i\int_0^t d\tau
g(\tau)e^{i\varphi_{\mu}(\tau)}\right]e^{-i\varphi_{\mu}(t)}$ and $\varphi_\mu(t)=\int_0^t d\tau [\omega_0+2\mu(t)]$). In the lowest ("classical") approximation when only the term $\sim \hbar$ from (\ref{JR}) is kept, the $\nu$-integration results in the $\delta$ function $\prod_{\tau}\delta\left[\mu(\tau)-|a_{\mu}(\tau)|^2\right]$, so that $\mu(t)$ coincides with the classical action $I_c(t)$ [see Eq.~(\ref{I_c})]. The only contribution comes then from the periodic classical orbit which passes through the phase point ${\overset{\circ}\alpha}$. The corresponding fidelity amplitude is simply equal to $f_{\overset{\circ}\alpha}(t)=
\exp\left[i\sigma\int_0^t d\tau I_c(\tau)\right]$.

The first correction, given by the term $\sim \hbar^2$ in the functional ${\cal R}$, describes the quantum fluctuations. The functional integration still can be carried out exactly \cite{Sokolov84}. Now a bunch of trajectories contributes, which satisfy the equation $\mu(t;\delta)=|\delta+a_{\mu}(t)|^2-|\delta|^2$ for all ${\delta}$ within a quantum cell $\sim\hbar$. This equation can still be written in the form of the classical equation (\ref{I_c}) if we define the classical action along a given trajectory as ${\tilde
I}_c(t)=|a_\mu(t)+\delta|^2=\mu(t;\delta)+|\delta|^2=
I_c\left(\omega_0-2|\delta|^2;{\overset{\circ}\alpha}^*
+\delta^*,{\overset{\circ}\alpha}+\delta;t\right)$. For any given $\delta$ this equation describes the classical action of a nonlinear oscillator with linear frequency $\omega_0-2|\delta|^2$, which evolves along a classical trajectory starting from the point ${\overset{\circ}\alpha}+\delta$. One then obtains (up to the irrelevant overall phase factor $e^{-i\omega_0 t/2\hbar}$)
\begin{equation}
f_{\overset{\circ}\alpha}(t)=\frac{2}{\pi\hbar}\int
d^2\delta
e^{-\frac{2}{\hbar}|\delta|^2}\exp\left\{i\frac{\sigma}{2}
\left[{\tilde\varphi_c(t)}-{\tilde\varphi_c(0)}\right]\right\},
\label{fidocoherent}
\end{equation}
where the "classical" phase
${\tilde\varphi_c(t)}=\varphi_c(\omega_0-
2|\delta|^2;\overset{\circ}\alpha^*+\delta^*,
\overset{\circ}\alpha+\delta;t)= \int_0^t
d\tau\left[\omega_0-2|\delta|^2+2{\tilde I}_c(\tau)\right]$. This
expression gives the fidelity amplitude in  the "initial value
representation" \cite{Miller01,Vanicek03}. We stress that the
fidelity $F_{\overset{\circ}\alpha}=|f_{\overset{\circ}\alpha}|^2$
does not decay in time if the quantum fluctuations described by the
integral over $\delta$ in (\ref{fidocoherent}) are neglected.

On the initial stage of the evolution, while the phases
${\tilde\varphi_c(t)}$ are not yet randomized and still remember the
initial conditions, we can expand ${\tilde\varphi_c(t)}$ over the
small shifts $\delta$. Keeping only linear and quadratic terms in
(\ref{fidocoherent}) we get after double Gaussian integration
\begin{equation}\label{Sexp}
F_{\overset{\circ}\alpha}(t)=\left[1+
\left(\frac{\varepsilon}{2}\right)^2\left(\frac{\partial \varphi_c(t)}{\partial\omega_0}\right)^2\right]^{-1}
\exp\left\{-\frac{\varepsilon^2}{4\hbar}\Big|\frac{\partial \varphi_c(t)}{\partial\overset{\circ}\alpha}\Big|^2\left[1+
\left(\frac{\varepsilon}{2}\right)^2\left(\frac{\partial \varphi_c(t)}{\partial\omega_0}\right)^2\right]^{-1}\right\}\,.
\end{equation}
Due to exponential local instability of the classical dynamics  the
derivatives $\Big|\frac{\partial
\varphi_c(t)}{\partial\overset{\circ}\alpha}\Big|,\,\Big|\frac{\partial
\varphi_c(t)}{\partial\omega_0}\Big|\propto e^{\Lambda t}$ where
$\Lambda$ is the Lyapunov exponent. So, the function (\ref{Sexp}),
up to  time $t\ll\frac{1}{\Lambda}\ln\frac{2}{\varepsilon}$, decays
superexponentially:
$F_{\overset{\circ}\alpha}(t)\approx\exp\left(-\frac{\varepsilon^2}
{4\hbar}\Big|\frac{\partial
\varphi_c(t)}{\partial\overset{\circ}\alpha}\Big|^2\right)=
\exp\left(-\frac{\varepsilon^2}{4\hbar}e^{\Lambda t}\right)$
\cite{Silvestrov02, Iomin04}. During this time the contribution of
the averaging over the initial Gaussian distribution in the
classical $\overset{\circ}\alpha$ phase plane dominates while the
influence of the quantum fluctuations of the linear frequency
described by the $\omega_0$-derivative remains negligible. Such a
decay has, basically, a classical nature \cite{eckhardt} and the
Planck's constant appears only as the size of the initial
distribution. On the contrary for larger times the quantum
fluctuations of the frequency control the fidelity decay, which
becomes exponential: $F_{\overset{\circ}\alpha}(t)\propto
\left(\frac{\partial
\varphi_c(t)}{\partial\omega_0}\right)^{-2}=\exp(-2\Lambda t)$.

\section{Fidelity for mixed states versus classical correlation functions}
\label{sec:fidmixed} Now we discuss the decay of the fidelity ${\cal
F}(t)$ when the initial condition corresponds to a broad incoherent
mixture. More precisely, we consider a mixed initial state
represented by a Glauber's diagonal expansion \cite{Glauber63}
$\overset{\circ}\rho=~\int d^2\overset{\circ}\alpha {\cal
P}(|\overset{\circ}\alpha-
\overset{\circ}\alpha_c|^2)|\overset{\circ}\alpha\rangle
\langle\overset{\circ}\alpha|$ with a wide positive definite weight
function ${\cal P}$ which covers a large number of quantum cells.
Then our fidelity, defined as in Eq.~(\ref{F1}), equals ${\cal
F}(t;{\overset{\circ}\alpha_c})=|f(t;{\overset{\circ}\alpha_c})|^2$,
where
\begin{equation}\label{ampmx}
\begin{array}{c}
f(t; {\overset{\circ}\alpha_c})\equiv\int
d^2\overset{\circ}\alpha
{\cal P}(|\overset{\circ}\alpha-\overset{\circ}\alpha_c|^2)
f_{\overset{\circ}\alpha}(t)\\
\approx \frac{2}{\pi\hbar}\int d^2\delta
e^{-\frac{2}{\hbar}|\delta|^2} \int
d^2\overset{\circ}\alpha {\cal
P}(|\overset{\circ}\alpha -(\overset{\circ}\alpha_c+\delta)|^2)e^{i\frac{\sigma}{2}
\left[\overline{\varphi}_c(t)-\overline{\varphi}_c(0)\right]},
\end{array}
\end{equation}
with $\overline{\varphi}_c(t)=\varphi_c(\omega_0-2|\delta|^2;
\overset{\circ}\alpha^*,\overset{\circ}\alpha; t)$. The inner
integral over $\overset{\circ}\alpha$ looks like a classical
correlation function. In the regime of classically chaotic motion
this correlator cannot appreciably depend on either the exact
location of the initial distribution in the classical phase space or
on small variations of the value of the linear frequency. Indeed,
though an individual classical trajectory is exponentially sensitive
to variations of initial conditions and system parameters, the
manifold of all trajectories which contribute to (\ref{ampmx}) is
stable \cite{Cerruti02}. Therefore, we can fully disregard the
$\delta$-dependence of the integrand, thus obtaining
\begin{equation}\label{Clfmix}
f(t; {\overset{\circ}\alpha_c})\approx\int
d^2\overset{\circ}\alpha {\cal
P}(|\overset{\circ}\alpha-\overset{\circ}\alpha_c|^2)\exp\left\{i\frac{\sigma}{2}
\left[\varphi_c(t)-\varphi_c(0)\right]\right\}.
\end{equation}
This is the main result of our paper which directly relates
the decay of \textit{quantum} fidelity to that of correlation functions of \textit{classical} phases (see Eq.~(\ref{corrphy_c})). No quantum feature is present in the r.h.s. of (\ref{Clfmix}).

The decay pattern of the function ${\cal F}(t)=|f(t; {\overset{\circ}\alpha_c})|^2$ depends on the value of the parameter $\sigma=\varepsilon/\hbar$. In particular, for $\sigma\ll 1$, we recover the well known Fermi Golden Rule (FGR) regime. Indeed, in this case the cumulant expansion can be used, $\ln f(t; {\overset{\circ}\alpha_c})=\sum_{\kappa=1}^{\infty} \frac{(i\sigma)^{\kappa}}{\kappa!}\chi_{\kappa}(t)\,.$ All the cumulants are real, hence, only the even ones are significant. The lowest of them
\begin{equation}\label{qum}
\begin{array}{c}
\chi_2(t)=\int_0^t d\tau_1\int_0^t d\tau_2\langle \left[
I_c(\tau_1)- \langle I_c(\tau_1)\rangle\right]
\left[ I_c(\tau_2)-\langle I_c(\tau_2)
\rangle\right]\rangle\equiv\int_0^t d\tau_1 \int_0^t
d\tau_2 K_I(\tau_1,\tau_2)\;.
\end{array}
\end{equation}
is positive. Assuming that the classical autocorrelation function decays exponentially, $K_I(\tau_1,\tau_2)=\langle\left(\Delta I_c\right)^2\rangle\exp\left(-|\tau_1-\tau_2|/\tau_I\right)$ with some
characteristic time $\tau_I$, we obtain $\chi_2(t)=2\langle\left(\Delta
I_c\right)^2\rangle\tau_It=2Kt$ for the times $t>\tau_I$ and arrive, finally, at the FGR decay law ${\cal F}(t; {\overset{\circ}\alpha}_c) =\exp(-2\sigma^2Kt)$ \cite{Cerruti02,Jacquod01,Prosen02}. Here $K=\int_0^{\infty}d\tau K_I(\tau,0)= \langle\left(\Delta I_c\right)^2\rangle\tau_I\,.$

The significance of the higher connected correlators
$\chi_{\kappa\geq 4}(t)$ grows with the increase of the parameter
$\sigma$. When this parameter roughly exceeds one, then the cumulant
expansion fails and the FGR approximation is no longer valid. In the
regime $\sigma\gtrsim 1$, the decay rate of the function ${\cal
F}(t; {\overset{\circ}\alpha}_c)=\big|f(t;
{\overset{\circ}\alpha_c})\big|^2$ ceases to depend on $\sigma$
\cite{Zaslavsky88} and coincides with the decay rate $1/\tau_c$ of
the correlation function (\ref{corrphy_c}),
\begin{equation}\label{Ldecay}
{\cal F}(t; {\overset{\circ}\alpha}_c)=\exp(-t/\tau_c)\,.
\end{equation}
This rate is intimately related to the local instability of the chaotic classical motion though it is not necessarily given by the Lyapunov exponent $\Lambda$ (it is worth noting in this connection that the Lyapunov exponent diverges in our driven nonlinear oscillator model).

Returning to the averaged fidelity (\ref{incoh}), it can be decomposed
into the sum of a mean ($\left|\overline{f}\right|^2\equiv {\cal F}$) and a fluctuating part:
\begin{equation}\label{Faver}
\overline{F(t)}= {\cal F}(t)+\overline{\left|f(t)- \overline{f(t)}\right|^2}\,.
\end{equation}
As we have already stressed above, the two fidelities ${\cal F}$ and
$\overline{F}$ are quite different in nature. Nevertheless, due to
the dephasing induced by classical chaos, the decays of these two
quantities are tightly connected: they both decay with the same rate
though the decay of $\overline{F(t)}$ is delayed by a time $t_d$. To
show this, let us make use of the Fourier transform of the fidelity
operator:
\begin{equation}\label{Fourf}
{\hat f}(t)=\frac{1}{\pi} \int d^2\eta\, q(\eta^*,\eta;t)
{\hat D}(\eta), \quad q(\eta^*,\eta;t)=Tr\left[{\hat D}^{\dag}(\eta){\hat f}(t)\right],
\end{equation}
where ${\hat D}(\eta)=\exp(\eta a^{\dag}-\eta^* a)$ is the displacement operator of coherent states. The Fourier transform $q$ satisfies the obvious initial condition $q(\eta^*,\eta;0)=\pi\delta^{(2)}(\eta)$.
On the other hand, unitarity of the fidelity operator yields
\begin{equation}\label{Unitf}
\frac{1}{\pi} \int d^2 \kappa\, e^{\frac{1}{2}(\omega\kappa^*-\omega^*\kappa)}\,Q(\omega,\kappa;t)\,
=\pi\delta^{(2)}(\omega).
\end{equation}
Here the shorthand
\begin{equation}\label{Srthnd}
Q(\omega,\kappa;t)\equiv q^*\left(\kappa^*-\frac{1}{2}\omega^*,\kappa-\frac{1}{2}\omega;t\right)
q\left(\kappa^*+\frac{1}{2}\omega^*,\kappa+\frac{1}{2}\omega;t\right)
\end{equation}
has been used. The function $Q$ factorizes at the initial moment $t=0$ as
$Q(\omega,\kappa;0)=\pi^2\,\delta^{(2)}(\kappa)\,\delta^{(2)}(\omega)$.

>From eq. (\ref{F1}, \ref{Fourf}) we obtain
\begin{equation}\label{FcalF}
{\cal F}(t)=\frac{1}{\pi^2}\int d^2 \omega\,e^{-\left(\frac{\Delta}{2\hbar}
+\frac{1}{4}\right)|\omega|^2}\,
\int d^2\kappa\, e^{-\left(\frac{2\Delta}{\hbar}
+1\right)|\kappa|^2}\,Q(\omega,\kappa;t)
\end{equation}
and
\begin{equation}\label{FaverF}
\overline{F(t)}=\frac{1}{\pi^2}\int d^2 \omega\,e^{-\left(\frac{\Delta}{\hbar}
+\frac{1}{4}\right)|\omega|^2}\,
\int d^2\kappa\, e^{-|\kappa|^2}\,Q(\omega,\kappa;t)\,.
\end{equation}
Since we have assumed that the width of the initial mixture
$\Delta\gg \hbar$, the essential difference between the two latter
expressions lies in the $\kappa$-integration domain which is
determined by the exponential factor. However, at the initial moment
$t=0$ this difference is not relevant since the function $Q$ is
sharply peaked. Then, in the evolution process the function $Q$
widens so that the exponential factors begin to define the
integration range. This effect, in Eq.~(\ref{FcalF}), takes place
almost from the very beginning and therefore, after a very short
time,
\begin{equation}\label{apprcalF}
{\cal F}(t)\approx \left(\frac{\hbar}{\Delta}\right)^2\,
\Big|q(0,0;t)\Big|^2\approx \exp\left(-t/\tau_c\right)\,.
\end{equation}
(In the second equality we took into account the previously obtained
result (\ref{Ldecay})). On the contrary, the cut in the integration
over $\kappa$ is appreciably weaker in Eq.~(\ref{FaverF}). As long
as the factor $Q$ still decays faster than $e^{-|\kappa|^2}$ the
latter can be substituted by unity and the $\kappa$ integration
gives approximately $\pi\delta^{(2)}(\omega)$ as in the unitarity
condition (\ref{Unitf}). Up to this time $t_d$ the function
$\overline{F(t)}$ remains very close to one. When $t>t_d$,
\begin{equation}\label{appraverF}
\overline{F(t>t_d)}\approx \frac{\hbar}{\Delta}\,\Big|q(0,0;t>t_d)\Big|^2\approx \exp\left(-\frac{t-t_d}{\tau_c}\right)\,.
\end{equation}
Comparison of the two last equations allows us to estimate the delay time to be $t_d=\tau_c\ln\frac{\Delta}{\hbar}\,$. On the other hand, for the Peres' mixed-state fidelity (\ref{Fm}) we get
\begin{equation}\label{FFm}
F(t)=\frac{1}{\pi^2}\int d^2 \omega\,e^{-\left(\frac{\Delta}{2\hbar}
+\frac{1}{4}\right)|\omega|^2}\,
\int d^2\kappa\, e^{-\frac{\hbar}{2\Delta} |\kappa|^2}\, Q(\omega,\kappa;t)\,.
\end{equation}
When $\Delta\gg \hbar$ the $\kappa$ integration is not cut and the decay of $F(t)$, contrary to Eqs.~(\ref{apprcalF}, \ref{appraverF}), is determined by the large $\kappa$ asymptotic behavior of the function $q$. The fidelity (\ref{Fm}, \ref{FFm}) has a well defined classical limit which coincides with the classical fidelity \cite{Prosen02,eckhardt,Benenti03} and decays due to the phase flow out of the phase volume initially occupied. This has nothing to do with dephasing. In particular, if the initial distribution is uniform in the whole phase space the fidelity (\ref{Fm}) never decays.

In closing this section, we discuss possible choices of the
initial mixture ${\cal P}$.
If we choose $\overset{\circ}\alpha_c=0$, the initial mixed state reads as follows in the eigenbasis of the Hamiltonian ${H}^{(0)}=\hbar\omega_0 n+\hbar^2 n^2$ of the autonomous oscillations:
\begin{equation}\label{inrho}
\overset{\circ}\rho=\int d^2\overset{\circ}\alpha\, {\cal P}(|\overset{\circ}\alpha|^2)|\overset{\circ}\alpha\rangle
\langle\overset{\circ}\alpha|=\sum_{n=0}^{\infty}\,
\overset{\circ}{\rho}_n\,|n\rangle\langle n|,
\end{equation}
where
\begin{equation}\label{inrhon}
\overset{\circ}{\rho}_n=\frac{\pi}{n!}\int_0^{\infty} d\overset{\circ}I\,
{\cal P}(\overset{\circ}I)\,
e^{-\overset{\circ}I/\hbar}
\left(\overset{\circ}I/\hbar\right)^n, \,\,\,\, \overset{\circ}I=
|\overset{\circ}\alpha|^2\,.
\end{equation}
This formula is inverted as
\begin{equation}\label{inrhonin}
{\cal P}(\overset{\circ}I)=\frac{e^{\overset{\circ}I/\hbar}}{2\pi^2}\,
\hbar\int_{-\infty}^{\infty}dk\, e^{ik\overset{\circ}I}\,R(k), \quad
R(k)=\sum_{n=0}^{\infty} \overset{\circ}{\rho}_n\,(-i\hbar k)^n\,.
\end{equation}
Therefore, our initial state is a totally incoherent mixture of the eigenstates $|n\rangle$. In particular, it can be the thermal distribution $\overset{\circ}{\rho}_n=\exp\left(-E^{(0)}/T\right)$, a choice
of particular interest for experimental investigations.

\section{Fidelity decay in optical lattices: the kicked rotor model}
\label{sec:krot} 
\begin{figure}
\centerline{\epsfxsize=8.cm\epsffile{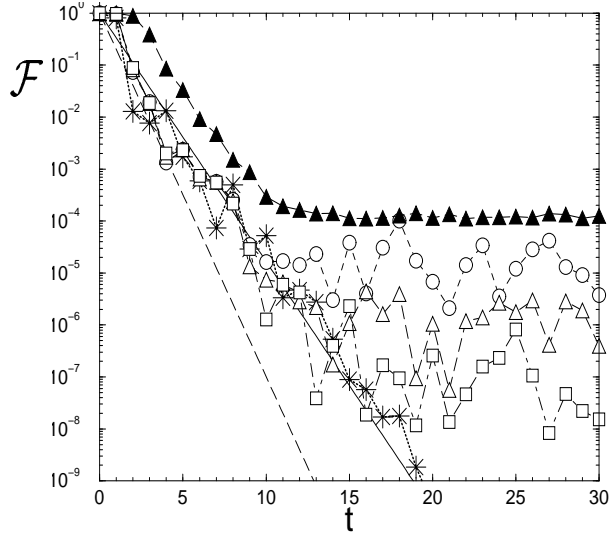}}
\caption{Decay of the fidelity ${\cal F}$ for the kicked rotor model with $K=10$, perturbation strength $\epsilon/\hbar=1.1$, $\hbar=3.1\times 10^{-3}$ (circles), $7.7\times 10^{-4}$ (empty triangles), and $1.9\times 10^{-4}$ (squares). Full triangles show the average fidelity $\overline{F}$ for $\hbar=7.7\times 10^{-4}$. Stars give the decay of the classical angular correlation function. The straight lines denote exponential decay with rates given by the Lyapunov exponent $\Lambda\approx \ln(K/2)=1.61$ (dashed line) and by the exponent $\Lambda_1\equiv\tau_c^{-1}=1.1$ \cite{Silvestrov02} (solid line).}
\label{fig1}
\end{figure}

As a second example, we consider the kicked rotor
model \cite{izrailev}, described by the Hamiltonian
$H=\frac{p^2}{2}+K\cos\theta\sum_m\delta(t-m)$, with
$[p,\theta]=-i\hbar$. The kicked rotor has been experimentally
implemented by cold atoms in a standing wave of light
\cite{Moore,Ammann,Delande,dArcy}. Moreover, the fidelity amplitude
for this model can be measured if one exploits atom interferometry
\cite{zoller,darcy,raizenschleich,raizenseligman}. The classical
limit corresponds to the effective Planck constant $\hbar\to 0$. We
consider this model on the torus, $0\le \theta <2\pi$, $-\pi\le p
<\pi$. The fidelity ${\cal F}$ is computed for a static perturbation
$\epsilon p^2/2$, the initial state being a mixture of Gaussian wave
packets uniformly distributed in the region $0.2\le \theta/2\pi \le
0.3$, $0.3\le p/2\pi \le 0.4$. In Fig.~\ref{fig1} we show the decay
of ${\cal F}(t)$ in the semiclassical regime $\hbar\ll 1$ and for a
quantally strong perturbation $\epsilon/\hbar\sim 1$. It is clearly
seen that the fidelity ${\cal F}$ follows the decay of the classical
angular correlation function
$|\langle\exp\{i\gamma[\theta(t)-\theta(0)]\}\rangle|^2$ (with the
fitting constant $\gamma=2$).
 We remark that ${\cal F}$ decays with a rate
$\Lambda_1$ different from the Lyapunov exponent. We also show the
fidelity $\overline{F(t)}$, (see eq.(\ref{Faver})), averaged over
the pure Gaussian states building the initial mixture.

Note that the expected saturation values of $\overline{F}$ and
${\cal F}$ are $1/N$ and $1/(NM)$, respectively, where $N$ is the
number of states in the Hilbert space and $M$ is the number of
quantum cells inside the area $\Delta$. This expectation is a
consequence of the randomization of phases of fidelity amplitudes
and is borne out by the numerical data shown in Fig.~\ref{fig1}.

\section{Conclusions}
\label{sec:conclusions} In this paper we have demonstrated that the
decay of the quantum fidelity ${\cal F}$, when the initial state is
a fully incoherent mixture, is determined by the decay of a
classical correlation function, which is totally unrelated to
quantum phases. We point out that the classical autocorrelation
function in Eq.~(\ref{Clfmix}) reproduces not only the slope but
also the overall decay of the function ${\cal F}$. The classical
dynamical variable that appears in this autocorrelation function
depends on the form of the perturbation. Therefore the echo decay,
even in a classically chaotic system in the semiclassical regime
and with quantally strong perturbations, is to some extent
perturbation-dependent. The quantum dephasing described in this
paper is a consequence of internal dynamical chaos and takes place
in absence of any external environment. We may therefore conclude
that the underlying internal dynamical chaos produces a dephasing
effect similar to the decoherence due to the environment.

\section{Acknowledgements}
We are grateful to Toma\v z Prosen and Dima Shepelyansky for useful discussions. This work was supported in part by EU (IST-FET-EDIQIP), NSA-ARDA (ARO contract No. DAAD19-02-1-0086) and
the MIUR-PRIN 2005 ``Quantum computation with trapped particle arrays, neutral and
charged''.
V.S. acknowledges financial support from the RAS Joint scientific program "Nonlinear dynamics and
solitons".

\end{document}